\documentclass[12pt]{article}
\usepackage{amscd,amsmath,amssymb,amstext,amsthm,exscale,latexsym}
\usepackage{array,graphicx,bbm,dsfont,exscale,hhline}
\newcommand {\dl}   {\delta}       
        
\newcommand {\ve}   {\varepsilon}  
\newcommand {\lm}   {\lambda}

\newcommand {\vf }  {\varphi}      
         \newcommand {\om}  {\omega}
      \newcommand {\Om}  {\Omega}
\newcommand {\Th}   {\Theta}       
\newcommand {\pl}   {\partial}     \newcommand {\nb}  {\nabla}
       \renewcommand {\lim}{{\sf\,lim\,}}
\newcommand {\CE }  {{\cal E}}      
\newcommand {\sh}{{\sf\,sh\,}}         \newcommand   {\ex}{{\sf\,e}}

\newcommand {\MO}  {{\mathbb O}}   
   \newcommand {\MR}  {{\mathbb R}}
\newcommand {\MS}  {{\mathbb S}}   
\newcommand {\MU}  {{\mathbb U}}

   \newcommand {\Bb}  {\boldsymbol{b}}

\newcommand {\Go}  {\mathfrak{o}}   
   
\newcommand {\Gs}  {\mathfrak{s}}   
\newcommand {\Gu}  {\mathfrak{u}}

\begin{document}
\title     {The 't Hooft--Polyakov monopole in the geometric theory of defects}
\author    {M. O. Katanaev
            \thanks{E-mail: katanaev@mi.ras.ru}\\ \\
            \sl Steklov mathematical institute,\\
            \sl 119991, Moscow, ul.~Gubkina, 8}
\maketitle
\begin{abstract}
The 't Hooft--Polyakov monopole solution in Yang--Mills theory is given new
physical interpretation in the geometric theory of defects. It describes solids
with continuous distribution of dislocations and disclinations. The
corresponding densities of Burgers and Frank vectors are computed. It means that
the 't Hooft--Polyakov monopole can be seen, probably, in solids.
\end{abstract}
%******************************************************************************
\section{Introduction}
%*******************************************************************************
Important properties of real crystals such as plasticity, melting, growth,
etc., are mainly defined by defects of the crystalline structure which are
called dislocations. Moreover, many bodies posses a spin structure. For example,
ferromagnets are also characterized by the distribution of magnetic moments
described by the unit vector field. This unit vector field may also have defects
(singularities) which are called disclinations. Description of dislocations and
disclinations in elastic media is a very active field of research for more then
one century because of its importance for applications (see, e.g.,
\cite{LanLif70,Kosevi81}).

Real solids posses usually a crystalline structure and are often described by
models based on this crystalline structure especially at the quantum level. At
the same time, many properties of solids can be also described by the elasticity
theory in the continuous approximation. Discrete and continuous approaches
complement each other, and are both needed for our understanding of nature.

In this paper, we
consider only continuous approximation. In this approximation solids without
dislocations are described by the displacement vector field within the ordinary
elasticity theory. The spin structure of solids without disclinations is
described by the unit vector field ($n$-field) satisfying appropriate field
equations. In the presence of dislocations and disclinations there as a problem:
what variable are to be used? For example, real solids posses many defects, and
if we want to use continuous approximation for defect distributions then the
displacement vector field and $n$-filed do not exist because they are singular
at each point. The geometric theory of defects is aimed to resolve this problem.

The idea of geometric theory of defects is simple. In the continuous
approximation, a crystal with a spin structure is considered as elastic media
(manifold) with a given
metric and affine connection with torsion (the Riemann--Cartan geometry).
As usual, elastic deformations of media and distribution of the unit vector
field are described by the displacement and rotational angle vector fields. The
absence of defects means that displacement and unit vector fields are smooth. If
they are not continuous then we say that the media has defects. In general,
there are two types of defects: dislocations which are defects of elastic media
itself (discontinuity of the displacement vector field) and disclinations
corresponding to discontinuities of the unit vector field. If defects are
absent, then geometry is trivial: curvature and torsion are zero. In the
presence of defects, geometry becomes nontrivial. Dislocations give rise
to torsion and disclinations result in nontrivial curvature. The physical
meaning of torsion and curvature are surface densities of Burgers
\cite{Burger39A,Burger39B} and Frank \cite{Frank58} vectors, respectively,
\cite{KatVol92,Katana05}. The geometric theory of defects allows one to describe
single defects as well as their continuous distribution. For single defects,
torsion and curvature are zero everywhere except some points, lines or surfaces
where defects are located and where they have singularities. In the case of
continuous distribution of dislocations and disclinations, torsion and curvature
become nontrivial on the whole media, and instead of the displacement and
angular rotation field we use tetrad and $\MS\MO(3)$-connection as the
independent variables. The advantage is that these variables exist even in the
absence of the displacement and unite vector fields.

The history of geometric theory of defects goes back to 1950s [8--11]
\nocite{Kondo52,Nye53,BiBuSm55,Kroner58} when dislocations were related to
torsion for the first time. The review and earlier references can be found in
the book \cite{Kleine08}.

In the geometric approach to the theory of defects
\cite{KatVol92,Katana05,KatVol99}, we discuss
the model which is different from others in two respects. Firstly, we do not
have the displacement and unit vector fields as {\em independent} variables
because, in general, they are not continuous. Instead, the triad field and
$\MS\MO(3)$-connection are considered as the only independent variables. If
defects are absent, then the triad and $\MS\MO(3)$-connection reduce to partial
derivatives of the displacement and rotational angle vector fields (pure gauge
because torsion and curvature vanish).
In this case, the latter can be reconstructed. Secondly, the set of equilibrium
equations is different. We proposed the purely geometric set which coincides
with that of Euclidean three dimensional gravity with torsion. The nonlinear
elasticity equations and principal chiral $\MS\MO(3)$-model for the unit vector
field enter the model through the elastic and Lorentz gauge conditions
\cite{Katana03,Katana04,Katana05} which allow us to reconstruct the
displacement and unit vector fields in the absence of defects in full agreement
with classical models.

When a new model is proposed then one has to show how to obtain previous results
within new approach. A number of dislocations were described in the geometric
theory of defects and shown to be in agreement with the elasticity theory
\cite{Katana05}, which corresponds to linear approximation. Therefore the
geometric theory of defects does not contradict experimental data in the domain
where elasticity theory is valid. At the same time, the geometric theory of
defects have also different predictions, for example, for the deformation tensor
near the core of wedge dislocation. As far as we know, there is no experimental
confirmation or refutation of geometric theory of defects. So, the model is
still under theoretical development.

In this paper, we consider the possibility of physical interpretation of the
't Hooft--Polyakov monopole solution \cite{tHooft74,Polyak74} in the geometric
theory of defects.
The famous 't Hooft--Polyakov solution in the
$\MS\MU(2)$ gauge theory interacting with the triplet of scalar fields attracted
much interest in physics and mathematics (for review, see, for example,
\cite{ManSut04,Shnir05}). The solution is static and spherically symmetric.
Therefore, it reduces to minimization of three-dimensional Euclidean energy
expression which can be regarded as the free energy expression in solid state
physics. We consider the $\MS\MU(2)$-connection components as the
$\MS\MO(3)$-connection because their Lie algebras coincide, the triplet of
scalar fields being the source of defects. Moreover, we assume that the
$\MS\MO(3)$ group acts not in the isotopic space but in the tangent space to
space manifold $\MR^3$. The metric of the space remains
Euclidean. So the 't Hooft--Polyakov monopole corresponds to Euclidean vielbein
and nontrivial $\MS\MO(3)$-connection which give rise to nontrivial
Riemann--Cartan geometry of space.

So, the 't Hooft--Polyakov monopole solution has natural interpretation in solid
state physics describing elastic media with continuous distribution of
disclinations and dislocations. We compute the corresponding densities of Frank
and Burgers vectors.
%******************************************************************************
\section{Geometric theory of defects}
%*******************************************************************************
In this section we give short review of the geometric theory of defects and
introduce basic geometric notions: triad field and $\MS\MO(3)$-connection.
More details can be found in \cite{Katana05}.

We consider a three dimensional
continuous media described by a topologically trivial Riemann--Cartan manifold.
We use triad field $e_\mu{}^i$ and $\MS\MO(3)$-connection
$\om_\mu{}^{ij}=-\om_\mu{}^{ji}$, where Greek letters $\mu=1,2,3$ and Latin ones
$i,j=1,2,3$ denote world and tangent indices, respectively, as basic independent
variables. We assume that metric
$g_{\mu\nu}:=e_\mu{}^i e_\nu{}^j\dl_{ij}=\dl_{\mu\nu}$ is an
ordinary flat Euclidean metric, but connection is nontrivial and may have
singularities on some points, lines, or surfaces.

\begin{figure}[ht]%--------------------------------------------------
\hfill\includegraphics[width=.8\textwidth]{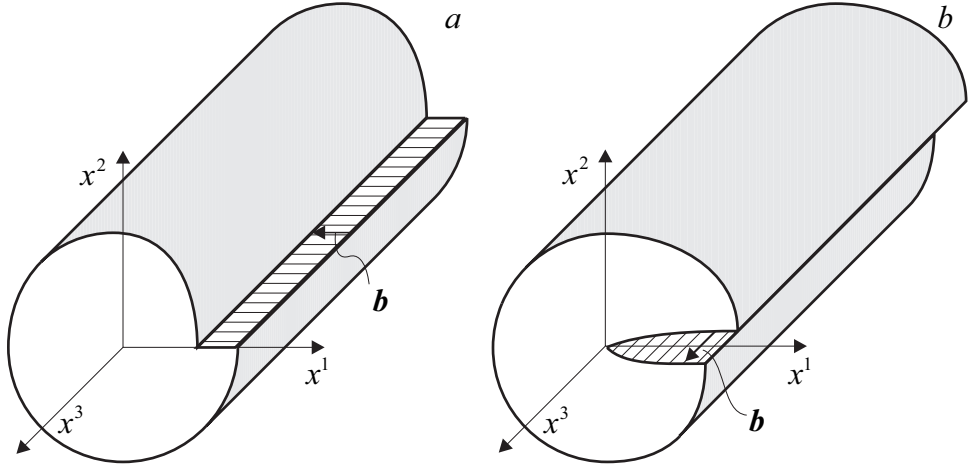}
\hfill {}
\\
\centering \caption{\label{fdislo} Straight linear dislocations.
            (\textit{a}) The edge
            dislocation. The Burgers vector $\Bb$ is perpendicular to the
            dislocation line. (\textit{b}) The screw dislocation. The Burgers
           vector $\Bb$ is parallel to the dislocation line.}
\end{figure}

The simplest and most widespread examples of linear dislocations are shown in
Fig.~\ref{fdislo} (see, e.g., \cite{LanLif70,Kosevi81}). They are produced as
follows.
We cut the medium along the half-plane $x^2=0$, $x^1>0$, move the upper part
of the medium located over the cut $x^2>0$, $x^1>0$ by the vector $\Bb$
towards the dislocation axis $x^3$, and glue the cutting surfaces.
The vector $\Bb$ is called the Burgers vector. In a general case, the
Burgers vector may not be constant on the cut. For the edge dislocation,
it varies from zero to some constant value $\Bb$ as it moves from the
dislocation axis. After the gluing, the media comes to the equilibrium state
called the edge dislocation, see Fig.~\ref{fdislo}\textit{a}.
If the Burgers vector is parallel to the dislocation line, it is called
the screw dislocation (Fig.~\ref{fdislo}\textit{b}).

%---------------------------------------------------------

From the topological standpoint, the medium containing several dislocations
or even the infinite number of them is still the Euclidean space $\MR^3$.
In contrast to the case of elastic deformations, the displacement vector
in the presence of dislocations is no longer a smooth function because
of the presence of cutting surfaces where it jumps.

The main idea of the geometric approach amounts to the following. To describe
single dislocations in the framework of elasticity theory, we must solve
equations for the displacement vector with some boundary conditions on the
cuts. This is possible for small number of dislocations. But, with an
increasing number of dislocations, the boundary conditions become so
complicated that the solution of the problem becomes unrealistic. Besides,
one and the same dislocation can be created by different cuts which
leads to an ambiguity in the displacement vector field. Another shortcoming
of this approach is that it cannot be applied to the description of a
continuous distribution of dislocations because the displacement
vector field does not exist in this case at all because it must have
discontinuities at every point. In the geometric approach, we consider the
triad field instead of the displacement vector field which is introduced as
follows.

Let a point of the medium has Cartesian coordinates $y^i$ in the ground
equilibrium state. After elastic deformation, this point has the coordinates
\begin{equation}                                                  \label{eeldef}
  y^i\mapsto x^i(y)=y^i+u^i(x),
\end{equation}
where $u^i(x)$ is the displacement vector field. We consider its components as
functions of final point position $x$.

In a general dislocation-present case, we do not have a preferred Cartesian
coordinate system in the equilibrium because there is no symmetry.
Therefore, we consider arbitrary global coordinates $x^\mu$, $\mu=1,2,3$, in
$\MR^3$. We use Greek letters for coordinates allowing
arbitrary coordinate changes. Then the Burgers vector for linear dislocation can
be expressed as the integral of the displacement vector
\begin{equation}                                        \label{eBurge}
  \oint_Cdx^\mu\pl_\mu u^i(x)=-\oint_Cdx^\mu\pl_\mu y^i(x)=-b^i,
\end{equation}
where $C$ is a closed contour surrounding the dislocation axis.
This integral is invariant under arbitrary coordinate
transformations $x^\mu\mapsto x^{\mu'}(x)$ and covariant under global
$\MS\MO(3)$-rotations of $y^i$. Here, components of the displacement vector
field $u^i(x)$ are considered with respect to the orthonormal basis in
the tangent space, $u=u^i e_i$. If components of the displacement vector
field are considered with respect to the coordinate basis $u=u^\mu\pl_\mu$,
the invariance of the integral (\ref{eBurge}) under general coordinate
changes is violated.

In the geometric approach, we introduce new independent variable -- the triad
-- instead of partial derivatives $\pl_\mu u^i$:
\begin{equation}                                                  \label{edevid}
  e_\mu{}^i(x):=\begin{cases} \pl_\mu y^i, &\text{outside the cut,}\\
               \lim\pl_\mu y^i, &\text{on the cut.}\end{cases}
\end{equation}
The triad is a smooth function on the cut by construction. We note that
if the vielbein was simply defined as partial derivatives $\pl_\mu y^i$,
then it would have the $\dl$-function singularity on the cut because functions
$y^i(x)$ have a jump. The Burgers vector can be expressed through the
integral over a surface $S$ having contour $C$ as the boundary:
\begin{equation}                                                  \label{eBurg2}
  \oint_Cdx^\mu e_\mu{}^i=\int\!\!\int_Sdx^\mu\wedge dx^\nu
  (\pl_\mu e_\nu{}^i-\pl_\nu e_\mu{}^i)=b^i,
\end{equation}
where $dx^\mu\wedge dx^\nu$ is the surface element. As a consequence of
the definition of the vielbein in (\ref{edevid}), the integrand is equal to
zero everywhere except at the dislocation axis. For the edge dislocation with
constant Burgers vector, the integrand has a $\dl$-function singularity at the
origin. The criterion for the presence of a dislocation is a violation of the
integrability conditions for the system of equations $\pl_\mu y^i=e_\mu{}^i$:
\begin{equation}                                        \label{eintco}
  \pl_\mu e_\nu{}^i-\pl_\nu e_\mu{}^i\ne0.
\end{equation}
If dislocations are absent, then the functions $y^i(x)$ exist and define
transformation to a Cartesian coordinates frame.

In the geometric theory of defects, the field $e_\mu{}^i$ is identified
with the triad. Next, we compare the integrand in (\ref{eBurg2}) with the
expression for the torsion in Cartan variables
\begin{equation}                                                  \label{ubbxvs}
  T_{\mu\nu}{}^i:=\pl_\mu e_\nu{}^i-\pl_\nu e_\mu{}^j-e_\mu{}^j\om_{\nu j}{}^i
  +e_\nu{}^j\om_{\mu j}{}^i.
\end{equation}
They differ
only by terms containing the $\MS\MO(3)$-connection $\om_{\mu j}{}^i$. This is
the ground for the introduction of the following postulate. In the geometric
theory of defects, the Burgers vector corresponding to a surface $S$ is defined
by the integral of the torsion tensor:
\begin{equation*}
  b^i:=\int\!\!\int_S dx^\mu\wedge dx^\nu T_{\mu\nu}{}^i.
\end{equation*}
This definition is invariant with respect to general coordinate transformations
of $x^\mu$ and covariant with respect to global rotations. Thus, the torsion
tensor has straightforward physical interpretation: it is equal to the surface
density of the Burgers vector.

If the curvature tensor for the $\MS\MO(3)$-connection
\begin{equation}                                                  \label{ubncbd}
  R_{\mu\nu}{}^{ij}:=\pl_\mu\om_\nu{}^{ij}-\pl_\nu\om_\mu{}^{ij}
  -\om_\mu{}^{ik}\om_{\nu k}{}^j+\om_\nu{}^{ik}\om_{\mu k}{}^j,
\end{equation}
is zero, then the connection is locally trivial, and there exists such
$\MS\MO(3)$ rotation that $\om_{\mu i}{}^j=0$. In this case, we return to
expression (\ref{eBurg2}).

\begin{figure}
\hfill\includegraphics[width=.75\textwidth]{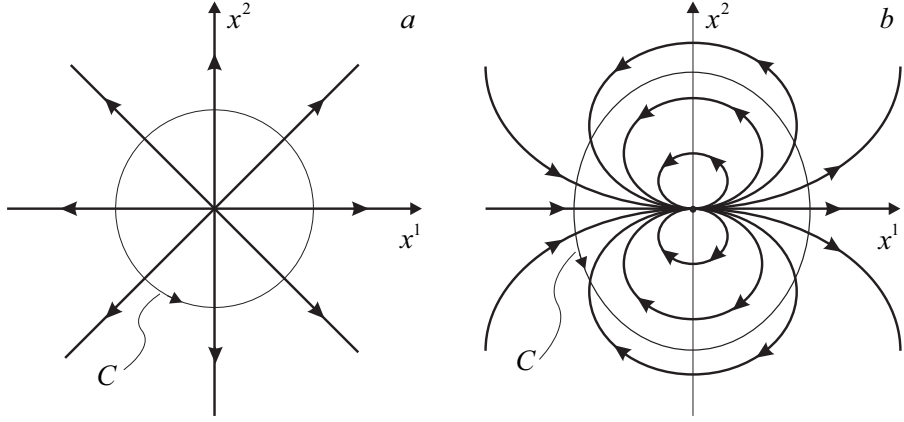}
\hfill {}
\centering\caption{Distribution of unit vector field in the $x:=x^1,y:=x^2$
plane for straight linear disclinations parallel to the $x^3$ axis,
for $|\Theta|=2\pi$ (\textit{a}) and $|\Theta|=4\pi$ (\textit{b}).}
\label{fdiscl}
\end{figure}

Next we give physical interpretation of the $\MS\MO(3)$-connection entering the
expression for torsion (\ref{ubbxvs}). To this end we consider more general
solids possessing spin structure, for example, ferromagnets or liquid crystals.
The spin structure is the unit vector field $n^i(x)$ $(n^in_i=1)$. It
can be described as follows. We fix some direction in the
medium $n_0^i$. Then the field $n^i(x)$ at a point $x$ can be uniquely defined
by the angular rotation field
$\theta^{ij}(x)=-\theta^{ji}(x) =\frac12\ve^{ijk} \theta_k$,  where
$\ve^{ijk}$ is the totally antisymmetric tensor and $\theta_k$ is a covector
directed along the rotation axis, its length being the rotation angle. Here and
in what follows, Latin tangent indices are raised and lowered with the help of
the flat Euclidean metric $\dl_{ij}$. So,
\begin{equation}                                                  \label{ubndhy}
  n^i=n_0^j S_j{}^i(\theta),
\end{equation}
where $S_j{}^i\in\MS\MO(3)$ is the rotation matrix corresponding to
$\theta^{ij}$ and parameterized as
\begin{equation}                                                  \label{elsogr}
  S_i{}^j=(e^{(\theta\ve)})_i{}^j=\cos\theta\,\dl_i^j
  +\frac{(\theta\ve)_i{}^j}\theta\sin\theta
  +\frac{\theta_i\theta^j}{\theta^2}(1-\cos\theta)\qquad \in\MS\MO(3)\, ,
\end{equation}
where $(\theta\ve)_i{}^j:=\theta_k\ve^k{}_i{}^j$ and
$\theta:=\sqrt{\theta^i\theta_i}$.
If the unit vector field is continuous then there are no disclinations.
Disclinations arise when the angular rotation field has discontinuities. The
simplest examples of linear disclinations are shown in
Fig.~\ref{fdiscl}, where the discontinuity of the angular rotation field occurs
on a half-plane cut from the $x^3$ axis to infinity, and the vector field $n$
lies in the perpendicular plane $(x^1,x^2)$.
%-------------------------------------------------------------------------------
%-------------------------------------------------------------------------------

A linear disclination is characterized by the Frank vector
\begin{equation}                                                  \label{etheta}
  \Th_i:=\frac12\ve_{ijk}\Th^{jk},
\end{equation}
where
\begin{equation}                                                  \label{eomega}
  \Th^{ij}:=\oint_Cdx^\mu\pl_\mu\theta^{ij},
\end{equation}
and the integral is taken along closed contour $C$ surrounding the disclination
axis. The length of the Frank vector is equal to the total angle of rotation of
the field $n^i$ as it goes around the disclination. For linear disclinations it
must be a multiple of $2\pi$. In the presence of
disclinations, the rotational angle field $\theta^{ij}(x)$ is no longer
continuous, and we must make some cuts for a given distribution of
disclinations and impose appropriate boundary conditions in order to define
$\theta^{ij}(x)$. In geometric theory of defects, instead of the rotational
angle field, we introduce the $\MS\MO(3)$-connection
\begin{equation}                                                  \label{edesoc}
  \om_\mu{}^{ij}:=\begin{cases}\pl_\mu\om^{ij}, &\text{outside the cut,}\\
                 \lim\pl_\mu\om^{ij}, &\text{on the cut.} \end{cases}
\end{equation}
in the way similar to the introduction of the triad field. Sure, we assume that
the limits on both sides of the cut exist and are equal. So the
$\MS\MO(3)$-connection is less singular then the rotational angle field by
definition.

Then the Frank vector for a surface $S$ is given by the integral of curvature
\begin{equation}                                                  \label{unbgtr}
  \Om^{ij}:=\iint_S dx^\mu\wedge dx^\nu R_{\mu\nu}{}^{ij}.
\end{equation}
If we have straight linear disclination with rotational symmetry, and vector
$n$ rotates in the perpendicular plane, then the $\MS\MO(3)$ group reduces to
abelian $\MS\MO(2)$ group, the nonlinear terms in the curvature (\ref{ubncbd})
disappear, and we return to the previous expression (\ref{eomega}) due to the
Stokes theorem.

The previous discussion refers to an isolated disclinations. If there is a
continuous distribution of disclinations the curvature differs from zero
everywhere, and the rotational angle field $\theta^{ij}$ does not exist.
Disclinations are said to be absent if and only if the curvature of
$\MS\MO(3)$-connection vanishes, $R_{\mu\nu i}{}^j=0$.
In this manner, the geometric theory of defects
describes single defects as well as their continuous distribution, in which
the phenomena of disclinations is replaced by the notion of curvature.
%******************************************************************************
\section{'t Hooft--Polyakov monopole}
%*******************************************************************************
Let us consider three-dimensional Euclidean space $\MR^3$ with Cartesian
coordinates $x^\mu$ and Euclidean metric $\dl_{\mu\nu}$, $\mu,\nu=1,2,3$,.
The spherically symmetric $\MS\MU(2)$-gauge fields $A_\mu{}^i$, $i=1,2,3$,
interacting with the triplet of scalar fields $\vf^i$ in the adjoint
representation minimize the three-dimensional energy \cite{ManSut04,Shnir05}
\begin{equation}                                                  \label{ubsghj}
  \CE:=\int \!d^3x\left(\frac14F^{\mu\nu i}F_{\mu\nu i}
  +\frac12\nb^\mu\vf^i\nb_\mu\vf_i+\frac14\lm\big(\vf^2-a^2\big)^2\right),
\end{equation}
where indices are raised and lowered by Euclidean metrics $\dl_{\mu\nu}$ and
$\dl_{ij}$,
\begin{equation}                                                  \label{uvvxfd}
\begin{split}
  F_{\mu\nu}{}^i:=&\pl_\mu A_\nu{}^i-\pl_\nu A_\mu{}^i
  +A_\mu{}^jA_\nu{}^k\ve_{jk}{}^i,
\\
  \nb_\mu\vf^i:=&\pl_\mu\vf^i+A_\mu{}^j\vf^k\ve_{jk}{}^i.
\end{split}
\end{equation}
-- are the curvature tensor components for $\MS\MU(2)$-connection and the
covariant derivative of scalar fields; $\lm>0, a>0$ -- are coupling constants,
$\ve_{ijk}$ is the totally antisymmetric tensor, $\ve_{123}:=1$, and
$\vf^2:=\vf^i\vf_i$.

The spherically symmetric ansatz is
\begin{equation}                                                  \label{uncbgf}
  A_\mu{}^i=\frac{\ve_\mu{}^{ij}x_j(K-1)}{r^2},\qquad\vf^i=\frac{x^i H}{r^2},
\end{equation}
where $K(r)$ and $H(r)$ are some dimensionless functions on radius
$r:=\sqrt{x^2}$.

The Euler--Lagrange equations for functional (\ref{ubsghj}) in the spherically
symmetric case reduce to
\begin{equation}                                                  \label{ubvxgd}
\begin{split}
  r^2K''=&K\big(K^2+H^2-1\big),
\\
  r^2H''=&2HK^2+\lm\left(H^2-a^2r^2\right)H.
\end{split}
\end{equation}
At present we know only one exact analytic solution to this system of equations
for $\lm=0$
\begin{equation}                                                  \label{uvxcse}
  K=\frac{ar}{\sh(ar)},\qquad H=\frac{ar}{\tanh(ar)}-1,
\end{equation}
which is called the Bogomol'nyi--Prasad--Sommerfield solution
 \cite{PraSom75,Bogomo76}. It is easily checked that this solution has finite
energy.

The Lie algebra $\Gs\Gu(2)$ is isomorphic to $\Gs\Go(3)$, and we can consider
energy (\ref{ubsghj}) as the three-dimensional Euclidean functional for
$\MS\MO(3)$-connection interacting with the triplet of scalar fields $\vf^i$ in
the fundamental representation. We assume, that this is the expression for the
free energy describing static distribution of disclinations and dislocations in
elastic media with defects, the triplet of scalar fields being the source of
defects.

The Euclidean metric means that elastic stresses are absent in media. The
Cartan variables for monopole solutions are
\begin{equation}                                                  \label{ubbcvl}
  e_\mu{}^i=\dl_\mu^i,\qquad\om_\mu{}^{ij}=A_\mu{}^k\ve_k{}^{ij}
  =(\dl_\mu^jx^i-\dl_\mu^ix^j)\frac{K-1}{r^2},
\end{equation}
where we use the spherically symmetric $\MS\MO(3)$-connection (\ref{uncbgf}).
The curvature and torsion are expressed through Cartan variables as usual by
Eqs.(\ref{ubbxvs}), (\ref{ubncbd}).
In the considered case, simple calculations yield the following expressions for
curvature and torsion:
\begin{align}                                                     \label{ubbcvh}
  R_{\mu\nu}{}^k:=\frac12R_{\mu\nu}{}^{ij}\ve_{ij}{}^k=F_{\mu\nu}{}^k
  =&\ve_{\mu\nu}{}^k\frac{K'}{r}-\frac{\ve_{\mu\nu}{}^jx_jx^k}{r^3}
  \left(K'-\frac{K^2-1}r\right),
\\                                                                \label{unbgtr}
  T_{\mu\nu}{}^k=&\left(\dl_\mu^kx_\nu-\dl_\nu^kx_\mu\right)\frac{K-1}{r^2}.
\end{align}

In the geometric theory of defects, curvature (\ref{ubbcvh}) and torsion
(\ref{unbgtr}) have physical meaning of surface densities of Frank and Burgers
vectors, respectively. That is they are equal to $k$-th components of respective
vectors on surface element $dx^\mu\wedge dx^\nu$. If $s^\mu$ is normal to the
surface element, then there are the following densities of Frank and Burgers
vectors:
\begin{align}                                                     \label{uvvxgd}
  f_\mu{}^i:=&\frac12\ve_\mu{}^{\nu\rho}R_{\nu\rho}{}^i
  =\frac1{3r}\dl_\mu^i\left(2K'+\frac{K^2-1}r\right)-\frac1{r}
  \left(\hat x_\mu\hat x^i-\frac13\dl_\mu^i\right)\left(K'-\frac{K^2-1}r\right),
\\                                                                \label{ubfgry}
  b_\mu{}^i:=&\frac12\ve_\mu{}^{\nu\rho}T_{\nu\rho}{}^i
  =\ve_\mu{}^{ij}\hat x_j\frac{K-1}{r},
\end{align}
where $\hat x^\mu:=x^\mu/r$ and tensor $f_\mu{}^i$ is decomposed into
irreducible components.

For the Bogomol'nyi--Prasad--Sommerfield solution functions $K(r)$ and $H(r)$
are given in Eq.\ (\ref{uvxcse}). They have the following asymptotics
\begin{equation}                                                  \label{unnvhf}
\begin{aligned}
  K\big|_{r\to0}\approx & 1-\frac{(ar)^2}6-\frac{(ar)^4}{120}, &
  \qquad K\big|_{r\to\infty}\approx & 2ar\ex^{-ar}\to0,
\\
  H\big|_{r\to0}\approx & 1+\frac{(ar)^2}3-\frac{2(ar)^4}{15},  &
  H\big|_{r\to\infty}\approx & ar-1\to\infty.
\end{aligned}
\end{equation}
The corresponding asymptotics of Frank and Burgers vector densities are
\begin{equation}                                                  \label{uvnfuy}
\begin{split}
  f_\mu{}^i\big|_{r\to0}\approx&-\frac13\dl_\mu^i\left(a^2+\frac7{90}a^4r^2
  \right)+\frac2{45}x_\mu x^i3a^4\to-\frac13\dl_\mu^ia^2,
\\
  b_\mu{}^i\big|_{r\to0}\approx & -\frac16\ve_\mu{}^{ij}x_j\left(a^2+
  \frac{a^4r^2}{20}\right)\to-\frac16\ve_\mu{}^{ij}x_ja^2,
\\
  \vf^i\big|_{r\to0}\approx & \frac13x^i\left(a^2-\frac{2a^4r^2}5\right)
  \to\frac13x^ia^2,
\\
  f_\mu{}^i\big|_{r\to\infty}\approx & -\frac{x_\mu x^i}{r^4}\to0,
\\
  b_\mu{}^i\big|_{r\to\infty}\approx & -\ve_\mu{}^{ij}x_j\frac1{r^2}\to0,
\\
  \vf^i\big|_{r\to\infty}\approx & \frac{x^i}r\left(a-\frac1{r}\right)
  \to\frac{x^i}ra.
\end{split}
\end{equation}
It implies, in particular, that the total energy (\ref{ubsghj}) is finite.
%******************************************************************************
\section{Conclusion}
%*******************************************************************************
The geometric theory of defects is aimed for description of dislocations and
disclinations in the continuous approximation. It is well suited for description
of single defects as well as their continuous approximation. In the present
paper, we consider media with Euclidean metric but nontrivial
$\MS\MO(3)$-connection. The 't Hooft--Polyakov monopole solution is the static
spherically symmetric solution of $\MS\MU(2)$ Yang--Mills theory. The
isomorphism of $\Gs\Gu(2)$ and $\Gs\Go(3)$ Lie algebras implies that the
't Hooft--Polyakov monopole may have new physical interpretation in solid state
physics. In contrast to the original model, the $\MS\MO(3)$ group acts now
not in the isotopic space but in the tangent space, giving rise to nontrivial
torsion and curvature. These geometrical notions have physical interpretation as
surface densities of Burgers and Frank vectors, respectively, in the geometric
theory of defects. These are explicitly computed for the
Bogomol'nyi--Prasad--Sommerfield solution. We are not aware what kind of media
is to be chosen for experimental observations and what kind of experiment can
be taken to confirm or disprove the geometric theory of defects but the mere
existence of such possibility seems to be interesting.

%\bibliography{2dgrav,3dgrav,book,defect,gravity,hamil,kalkle,math,my,qft}

\begin{thebibliography}{10}

\bibitem{LanLif70}
L.~D. Landau and E.~M. Lifshits.
\newblock {\em Theory of Elasticity}.
\newblock Pergamon, Oxford, 1970.

\bibitem{Kosevi81}
A.~M. Kosevich.
\newblock {\em Physical mechanics of real crystals}.
\newblock Naukova dumka, Kiev, 1981.
\newblock [in Russian].

\bibitem{Burger39A}
J.~M. Burgers.
\newblock {\em Proc.\ Kon.\ Ned.\ Akad.\ Wetenschap.}, 42:293--378, 1939.

\bibitem{Burger39B}
J.~M. Burgers.
\newblock {\em Proc.\ Kon.\ Ned.\ Akad.\ Wetenschap.}, 42:378--398, 1939.

\bibitem{Frank58}
F.~C. Frank.
\newblock On the theory of liquid crystals.
\newblock {\em Discussions Farad.\ Soc.}, 25:19--28, 1958.

\bibitem{KatVol92}
M.~O. Katanaev and I.~V. Volovich.
\newblock Theory of defects in solids and three-dimensional gravity.
\newblock {\em Ann.\ Phys.}, 216(1):1--28, 1992.

\bibitem{Katana05}
M.~O. Katanaev.
\newblock Geometric theory of defects.
\newblock {\em Physics -- Uspekhi}, 48(7):675--701, 2005.
\newblock https://arxiv.org/abs/cond-mat/0407469.

\bibitem{Kondo52}
K.~Kondo.
\newblock On the geometrical and physical foundations of the theory of
  yielding.
\newblock In {\em Proc. 2nd Japan Nat. Congr. Applied Mechanics}, pages 41--47,
  Tokyo, 1952.

\bibitem{Nye53}
J.~F. Nye.
\newblock Some geometrical relations in dislocated media.
\newblock {\em Acta Metallurgica}, 1:153, 1953.

\bibitem{BiBuSm55}
B.~A. Bilby, R.~Bullough, and E.~Smith.
\newblock Continuous distributions of dislocations: a new application of the
  methods of non-{R}iemannian geometry.
\newblock {\em Proc. Roy. Soc. London}, A231:263--273, 1955.

\bibitem{Kroner58}
E.~Kr\"oner.
\newblock {\em Kontinums Theories der Versetzungen und Eigenspanungen}.
\newblock Spriger--Verlag, Berlin -- Heidelberg, 1958.

\bibitem{Kleine08}
H.~Kleinert.
\newblock {\em Multivalued Fields in Condenced Matter, Electromagnetism, and
  Gravitation}.
\newblock World Scientific, Singapore, 2008.

\bibitem{KatVol99}
M.~O. Katanaev and I.~V. Volovich.
\newblock Scattering on dislocations and cosmic strings in the geometric theory
  of defects.
\newblock {\em Ann.\ Phys.}, 271:203--232, 1999.

\bibitem{Katana03}
M.~O. Katanaev.
\newblock Wedge dislocation in the geometric theory of defects.
\newblock {\em Theor.\ Math.\ Phys.}, 135(2):733--744, 2003.

\bibitem{Katana04}
M.~O. Katanaev.
\newblock One-dimensional topologically nontrivial solutions in the {S}kyrme
  model.
\newblock {\em Theor.\ Math.\ Phys.}, 138(2):163--176, 2004.

\bibitem{tHooft74}
G.~'t~Hooft.
\newblock Magnetic monopoles in unified gauge theories.
\newblock {\em Nucl.\ Phys.\ B}, 79(2):276--284, 1974.

\bibitem{Polyak74}
A.~M. Polyakov.
\newblock Particle spectrum in the quantum field theory.
\newblock {\em JETP Letters}, 20(6):194--195, 1974.

\bibitem{ManSut04}
N.~Manton and P.~Sutcliffe.
\newblock {\em Topological Solitons}.
\newblock Cambridge University Press, Cambridge, 2004.

\bibitem{Shnir05}
Ya. Shnir.
\newblock {\em Magnetic Monopoles}.
\newblock Springer--Verlag, Berlin, Heidelberg, 2005.

\bibitem{PraSom75}
M.~K. Prasad and C.~H. Sommerfield.
\newblock Exact classical solution for the 't hooft monopole and the julia-zee
  dyon.
\newblock {\em Phys.\ Rev.\ Lett.}, 35:760--762, 1975.

\bibitem{Bogomo76}
E.~B. Bogomol'nyi.
\newblock The stability of classical solutions.
\newblock {\em Sov.\ J.\ Nucl.\ Phys.}, 24(4):449, 1976.

\bibitem{Katana17C}
M.~O. Katanaev.
\newblock Chern--Simons term in the geometric theory of defects.
\newblock {\em Phys.\ Rev.\ D}, 96:84054, 2017.
\newblock https://doi.org/10.1103/PhysRevD.96.084054
  https://arxiv.org/abs/1705.07888 [gr-qc].

\end{thebibliography}
%\bibliographystyle{unsrt}

\end{document}